\UseRawInputEncoding
\documentclass[%
 reprint,
 superscriptaddress,
 amsmath,amssymb,
 aps,
]{revtex4-2}

\usepackage{xcolor}
\usepackage{graphicx}
\usepackage{dcolumn}
\usepackage{bm}
\usepackage[hidelinks,colorlinks=true,linkcolor=blue,urlcolor=blue,citecolor=blue]{hyperref}

\makeatletter
\newcommand{\mycaption}[2][]{%
  \begingroup%
  \renewcommand{\figurename}{\textbf{FIG.}}
  \renewcommand{\@caption@fignum@sep}{ \textbf{$\vert$} }%
  \renewcommand{\fnum@figure}{{\normalfont\bfseries \figurename~\thefigure}}
  \caption[#1]{#2}%
  \endgroup%
}
\makeatother

\makeatletter
\def\ignorecitefornumbering#1{%
     \begingroup
         \@fileswfalse
         #1
    \endgroup
}
\makeatother

\begin{document}


\title{Progress and prospects in the quantum anomalous Hall effect}
\author{Hang Chi}    
    \email{chihang@mit.edu}
    \affiliation{Francis Bitter Magnet Laboratory, Plasma Science and Fusion Center, Massachusetts Institute of Technology, Cambridge, Massachusetts 02139, USA}
    \affiliation{U.S. Army CCDC Army Research Laboratory, Adelphi, Maryland 20783, USA}
\author{Jagadeesh S. Moodera}    
    \email{moodera@mit.edu}
    \affiliation{Francis Bitter Magnet Laboratory, Plasma Science and Fusion Center, Massachusetts Institute of Technology, Cambridge, Massachusetts 02139, USA}
    \affiliation{Department of Physics, Massachusetts Institute of Technology, Cambridge, Massachusetts 02139, USA}

\date{\today}

\begin{abstract}
The quantum anomalous Hall effect refers to the quantization of Hall effect in the absence of applied magnetic field. The quantum anomalous Hall effect is of topological nature and well suited for field-free resistance metrology and low-power information processing utilizing dissipationless chiral edge transport. In this Perspective, we provide an overview of the recent achievements as well as the materials challenges and opportunities, pertaining to engineering intrinsic/interfacial magnetic coupling, that are expected to propel future development of the field. 
\end{abstract}

\maketitle

\begin{figure*} [bht]
\includegraphics{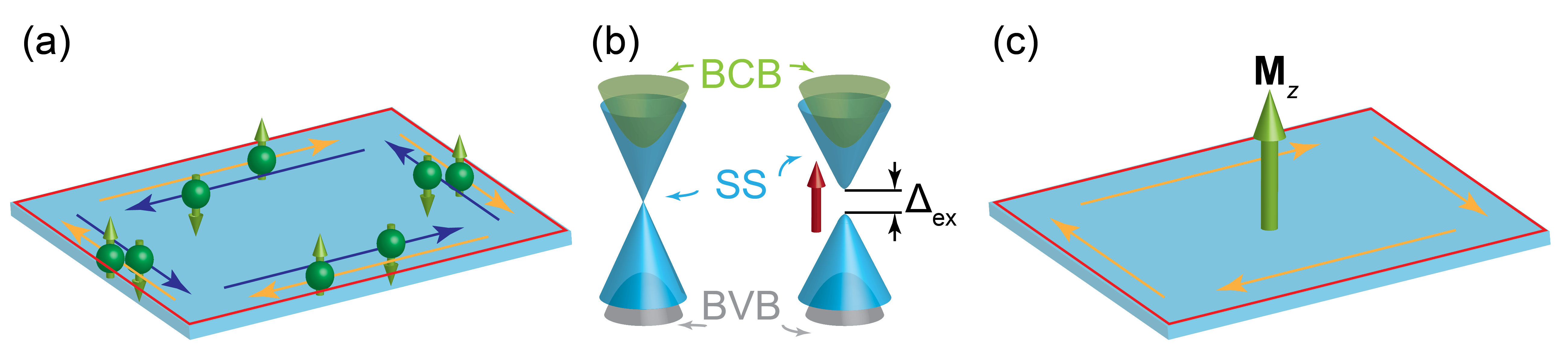}
\mycaption{\label{fig:fig1}\textbf{Quantum anomalous Hall effect with exchange gap opening in topological insulator films.} (a) The helical edge conduction in 2D TI protected by the time-reversal symmetry (TRS), displaying the quantum spin Hall effect. (b) Left, bulk conduction (BCB) and valence (BVB) bands of a typical 3D TI, along with the gapless surface states (SSs) protected by the TRS. Right, an exchange gap $\Delta\textsubscript{ex}$ induced in the SSs through broken TRS. (c) The chiral edge states in the quantum anomalous Hall effect with spontaneous perpendicular magnetization \textbf{M}$_z$ in magnetic TI. No external magnetic field is required for the dissipationless chiral edge current.}
\end{figure*}

\section{\label{sec:level1}Introduction}
The quantum anomalous Hall (QAH) effect represents one of the triumphs in conceptualizing topological aspects of electronic states in condensed matter physics \cite{RN1,RN2,RN2,RN3,RN4,RN5,RN6,RN7,RN8,RN9,RN10,RN11}. It constitutes an ever-evolving family of Hall effects \cite{RN12}. In 1879, Edwin Herbert Hall discovered that the Lorentz force leads to a transverse voltage when the longitudinal current in a conductor is subjected to a perpendicular external magnetic field \cite{RN13} -- an effect bears his name and inspires to push the scientific and technological frontiers \cite{RN14}. This ordinary Hall (OH) effect offers a direct tool assessing the charge carrier type and density in semiconductors as well as a practical probe measuring magnetic field. Hall later reported an unusual and stronger response in ferromagnets with qualitatively different field dependence \cite{RN15}. It hence came to be known as the anomalous Hall (AH) effect correlated with the spontaneous magnetization $M$. Its deep roots in topology and geometry were appreciated in recent times \cite{RN16}, enabled by the discoveries of integer \cite{RN17} and fractional \cite{RN18,RN19} quantum Hall (QH) effects in the 1980’s. 

For two-dimensional electron gas (2DEG) under strong applied magnetic field, the Hall resistance $R_{yx}$ develops well-defined plateaus quantized to exact value of $h/\nu e^2$ at which the longitudinal resistance $R_{xx}$ vanishes. Here $h$ is Planck's constant, $e$ is the elementary charge and $\nu$ is the filling factor that is topological in nature and corresponds to the Chern numbers (by integrating the Berry curvature \cite{RN20} over the first Brillouin zone) summed over occupied bands. In the integer QH regime, the quasi-1D chiral edge channels, each contributing a quantized Hall conductance $G_{xy} = e^2/h$, are immune to back scattering. For practical utilization of such dissipationless edge transport, e.g., in resistance metrology and low-energy cost electronics, it is natural to desire a QAH system displaying a quantized version of the AH effect. It allows QH states to prevail at zero magnetic field, circumventing the necessity of forming Landau levels as well as the often stringent requirement on sample mobility \cite{RN21,RN22,RN23}. 

It became clear, similar to the QH physics, the intrinsic AH conductivity $\sigma\textsubscript{AH}$ in a magnetic material is governed by the integral of the Berry curvature over occupied bands \cite{RN24,RN25,RN26,RN27,RN28,RN29}. Effort towards realizing the QAH effect was nonetheless at impasse for decades, as $\sigma\textsubscript{AH}$ is not quantized in metals with partially occupied bands and magnetic insulators with a non-zero Chern number are rare to come by. Since around 2005, however, new prospects emerged accompanying the discovery of topological insulators (TIs) with strong spin-orbit coupling (SOC) under the protection of time-reversal symmetry (TRS) \cite{RN30,RN31}. As shown in Fig.~\ref{fig:fig1}(a), a 2D TI characterized by a single $\mathbb{Z}_2$ topological invariant hosts a pair of spin-polarized helical edge states (can be roughly understood as two copies of the chiral QH states with opposite spin \cite{RN32}) that lead to the quantum spin Hall (QSH) effect with quantized six-terminal $R_{xx}$ of $h/2e^2$ and transverse spin-accumulation \cite{RN33,RN34,RN35,RN36,RN37,RN38,RN39}. 

Upon generalizing to 3D \cite{RN40}, as exemplified by the Bi$_2$Te$_3$ family of materials \cite{RN41,RN42,RN43,RN44}, TI features an insulating bulk and gapless helical surface states (SSs) with Dirac-like linear dispersion and spin-momentum locking, see Fig.~\ref{fig:fig1}(b). Introducing magnetic order to break the TRS in 2D TI, will intuitively lead to a QAH state – when one spin block is driven out of the topologically nontrivial band inverted regime into the normal one with vanishing $G_{xy}$ \cite{RN45}. Despite early progress, this route was not pursued further, largely owing to the finite magnetic field needed to induce quantization as a result of the paramagnetic nature of Mn doping in HgTe \cite{RN46}. As shown in Fig.~\ref{fig:fig1}(c), an alternative platform was brought forth to leverage 3D TI thin films instead, where each 2D SS of 3D TI contributes a half-quantized $G_{xy} = \pm e^2/2h$ under broken TRS \cite{RN47,RN48,RN49}. In 2013, this was realized in Cr-doped ternary (Bi,Sb)$_2$Te$_3$ thin films \cite{RN50}, soon gaining wide acceptance with ideal performance \cite{RN51,RN52,RN53,RN54,RN55,RN56,RN57,RN58,RN59,RN60,RN61}. More recently, the QAH effect was also demonstrated in one device of five-layer exfoliated intrinsic antiferromagnetic TI MnBi$_2$Te$_4$ \cite{RN62}, as well as in moir\'e materials, namely, magic-angle twisted bilayer graphene \cite{RN63} and AB-stacked MoTe$_2$/WSe$_2$ heterostructures \cite{RN64}, greatly expanding the available materials platforms and physical mechanisms for investigating the QAH phenomena.

\begin{figure*} [bht]
\includegraphics{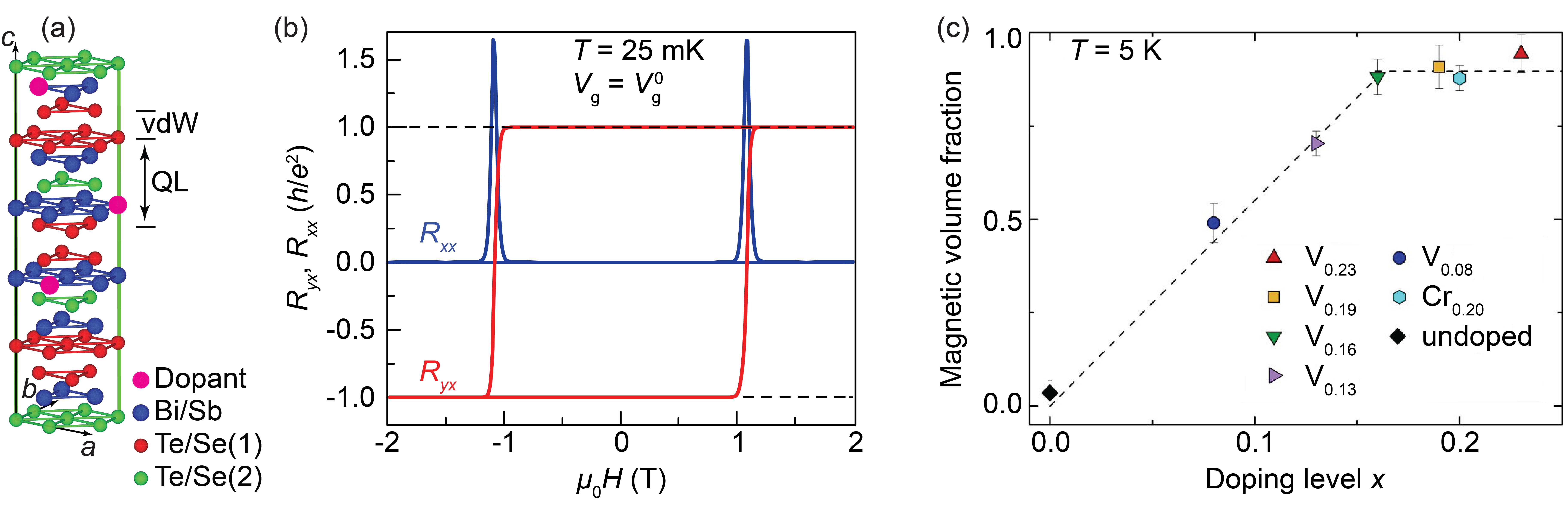}
\mycaption{\label{fig:fig2}\textbf{Quantum anomalous Hall effect in magnetically doped topolgoical insulator films.} (a) The crystal structure of (Bi,Sb)$_2$(Te,Se)$_3$ featuring quintuple layers (QLs) interconnected by weak van der Waals (vdW) bonding. Magnetic transition metal dopant V, Cr or Mn can occupy Bi/Sb site and induce long range magnetic ordering (while any dopant occupation inbetween layers would not be favorable). (b) The QAH state realized at charge neutrality point in a 4 QL thick V$_{0.11}$(Bi$_{0.29}$Sb$_{0.71}$)$_{1.89}$Te$_3$ film at $T$ = 25 mK. (c) The magnetic volume fraction measured by low energy muon spin rotation (LE-$\mu$SR) technique, as a function of the doping level $x$ in (V,Cr)$_x$(Bi,Sb)$_{2-x}$Te$_3$. Adapted with permissions from ref. \ignorecitefornumbering{\cite{RN56}}, Springer Nature Limited, (b); ref. \ignorecitefornumbering{\cite{RN87}}, American Physical Society, (c).}
\end{figure*}

\section{Realizing the QAH effect}

As shown in Fig.~\ref{fig:fig2}(a), the archetypical 3D TI of the Bi$_2$Te$_3$-based materials crystalize in a $R\bar{3}m$ ($D_{3d}^5$, No. 166) rhombohedral structure, featuring Te(1)-Bi-Te(2)-Bi-Te(1)-type quintuple layers (QLs) with weak van der Waals interlayer bonding \cite{RN65,RN66,RN67}. The topologically nontrivial members of the family display a single surface Dirac cone \cite{RN41} and band gap of 0.2--0.33 eV, 0.21--0.3 eV, and 0.13--0.2 eV, for Bi$_2$Se$_3$ \cite{RN68,RN69,RN70}, Sb$_2$Te$_3$ \cite{RN68,RN71,RN72}, and Bi$_2$Te$_3$ \cite{RN68,RN71,RN73,RN74,RN75}, respectively. The chemical stoichiometry and prevalence of defects sensitively affect the transport properties \cite{RN76}. When grown as thin films using molecular beam epitaxy (MBE), Bi$_2$Se$_3$ and Bi$_2$Te$_3$ are chiefly $n$ type owing to Se and Te vacancies, while Sb$_2$Te$_3$ is dominantly $p$ type due to Sb$_{\textrm{Te}}$ antisite defects. The isostructural nature of the compounds and availability of different carrier types offer the needed tunability of adjusting the chemical potential to zero-in on the surface transport.  

\begin{figure*} 
\includegraphics{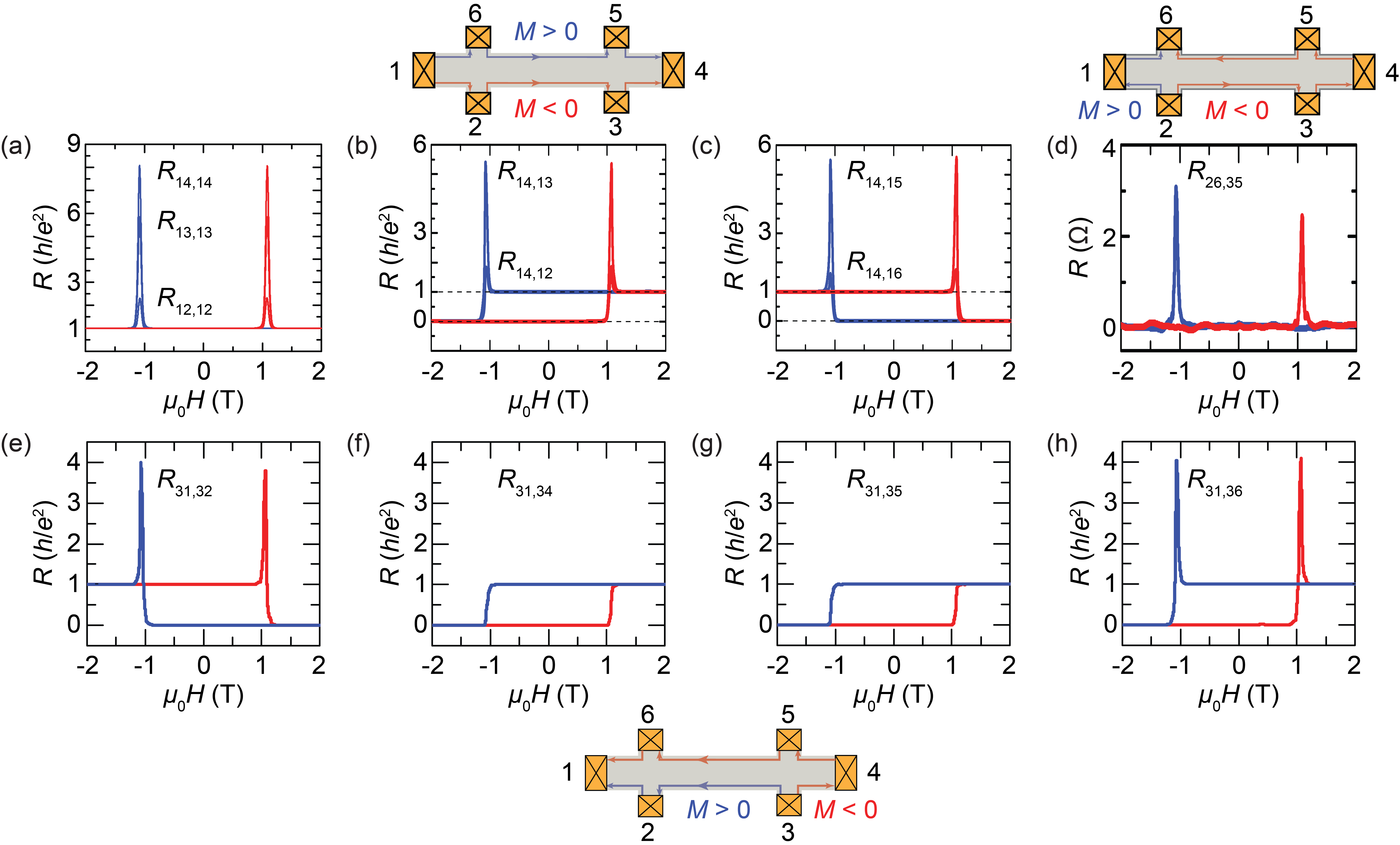}
\mycaption{\label{fig:fig3}\textbf{Local and nonlocal transport in the quantum anomalous Hall regime.} Magnetic field $\mu_{0}H$ dependence at $T$ = 25 mK for (a) two-terminal resistances $R_{12,12}$, $R_{13,13}$, $R_{14,14}$; three-terminal resistances (b) $R_{14,13}$, $R_{14,12}$; (c) $R_{14,15}$, $R_{14,16}$; (d) nonlocal resistance $R_{26,35}$; and additional three-terminal measurements (e) $R_{31,32}$; (f) $R_{31,34}$; (g) $R_{31,35}$; (h) $R_{31,36}$. The first (last) two subscripts in the resistance notation refer to the current (voltage) leads. The chiral current flows are depicted in the device schematics as insets: top left for (a-c), top right for (d) and bottom for (e-h). The red and blue lines indicate the chiral edge current for magnetization into ($M < 0$) and out of the plane ($M > 0$), respectively. Adapted with permission from ref. \ignorecitefornumbering{\cite{RN90}}, American Physical Society.}
\end{figure*}

Development of diluted magnetic semiconductors in the early 2000’s has provided important insights on how to induce long-range magnetic order in TIs. Taking Sb$_2$Te$_3$ for example, Mn doping in Mn$_x$Sb$_{2-x}$Te$_3$ bulk crystals does not stimulate long-range ordering for $x$ up to 0.045 \cite{RN77}, although ferromagnetism can be stabilized in Mn$_x$Bi$_{2-x}$Te$_3$ ($x$ up to 0.09) \cite{RN78} as well as topological crystalline insulator Mn$_x$Sn$_{1-x}$Te ($x$ up to 0.12) \cite{RN79}. Doping with V and Cr on the other hand is effective in introducing magnetic order in V$_x$Sb$_{2-x}$Te$_3$ ($0.01 \leq x \leq 0.03$) \cite{RN80} and Cr$_x$Sb$_{2-x}$Te$_3$ ($x$ up to 0.095) \cite{RN81}. With the incorporation of magnetic dopants (V/Cr) elevated to even higher levels in the out-of-equilibrium MBE growth environment, strong out-of-plane ferromagnetic ordering has been successfully demonstrated with impressively high Curie temperature $T\textsubscript{C}$ reaching 177 K and 190 K in V$_x$Sb$_{2-x}$Te$_3$ ($x$ up to 0.35) \cite{RN82} and Cr$_x$Sb$_{2-x}$Te$_3$ ($x$ up to 0.59) \cite{RN83}, respectively. These early works on relatively thick films (hundreds of nm) have benchmarked the solubility limit, while setting expectation of the achievable exchange energy in films with thickness relevant to the QAH effect, considering $T_{\textrm{C}}$ generally decreases upon reducing thickness \cite{RN84}. The high doping, however, is expected to weaken and eventually destroy the topological nature of the band due to reduced SOC. At an intermediate level, V$_x$Sb$_{2-x}$Te$_3$ ($x$ up to 0.10) thin films can sustain high surface mobility, leading to prominent Shubnikov-de Haas (SdH) quantum oscillations \cite{RN85}. 

Despite its larger band gap and more ideally positioned surface Dirac cone, the ferromagnetic response from Cr-doped Bi$_2$Se$_3$ is quite weak \cite{RN86}. The effort towards realizing the QAH effect has since largely focused on the ternary (Bi,Sb)$_2$Te$_3$ matrix, with systematically engineered thickness and Bi/Sb alloying ratio. The QAH state was first demonstrated by Cr doping in 5 QL Cr$_{0.15}$(Bi$_{0.1}$Sb$_{0.9}$)$_{1.85}$Te$_3$ \cite{RN50}. It was then discovered that V doping instead leads to a better reproducibility and a more robust QAH effect \cite{RN56}. As shown in Fig.~\ref{fig:fig2}(b), a nearly ideal QAH state is present at the charge neutrality point in 4 QL V$_{0.11}$(Bi$_{0.29}$Sb$_{0.71}$)$_{1.89}$Te$_3$, manifesting zero magnetic field $R_{yx}(0)$ = 1.00019 $\pm$ 0.00069 $h/e^2$, $R_{xx}(0)$ = 0.00013 $\pm$ 0.00007 $h/e^2$ and an AH angle $\alpha\textsubscript{AH}$ of 89.993$^\circ$ $\pm$ 0.004$^\circ$ at 25 mK. As revealed by low energy muon spin rotation (LE-$\mu$SR) spectroscopy in Fig.~\ref{fig:fig2}(c), the full magnetic volume fraction is achieved in (Cr,V)$_x$(Bi,Sb)$_{2-x}$Te$_3$ only at doping levels with $x \geq 0.16$. The evolution of effective magnetic ordering is consistent with formation and growth of ferromagnetic islands which eventually encompasses full volume upon cooling \cite{RN87}.

The edge current-voltage $I_{i} = (e^2/h)\sum_{j} (T_{ji} V_{i} - T_{ij} V_{j})$, is determined by the transmission probability $T_{ji}$ connecting the $i$th to the $j$th electrode in the Landauer-B\"uttiker theory \cite{RN88,RN89}. As shown in Fig.~\ref{fig:fig1}(c) and Fig.~\ref{fig:fig3}, in the QAH regime, the chiral edge modes propagate clockwise (counter-clockwise) for $M > 0$ ($M < 0$), leading to $T_{i,i+1} = 1$ ($T_{i+1,i} = 1$). The ideal dissipationless chiral edge transport at zero magnetic field has indeed been verified by comprehensive local and nonlocal magnetoresistance experiments \cite{RN52,RN90,RN91,RN92}.In quantum phase transitions concerning a QAH insulator, the temperature dependence of the derivative of $R_{xx}(H)$ at the critical field $H_{\textrm{c}}$ follows a power law scaling behavior $(dR_{xx}/dH)|_{H\textrm{c}} \propto T^{-\kappa}$, with critical exponent $\kappa$ in the range of 0.22 -- 0.62 \cite{RN93,RN94,RN95,RN96,RN97}. Utilizing a cryogenic current comparator, metrologically comprehensive low-current high-precision measurements of the QAH states have demonstrated $R_{yx}$ quantization within 1 part per million (ppm) of the von-Klitzing constant $R_{\textrm{K}}$ \cite{RN98,RN99,RN100}, and most recently down to 10 parts per billion (ppb) leveraging a permanent magnet design \cite{RN101}. It improves the prospects for zero-field quantum resistance standard and spintronics exploiting chiral edge transport, which favor a more robust QAH state at higher temperature, with even smaller $R_{xx}$ and larger breakdown current density \cite{RN102,RN103}. 

\section{Higher temperature QAH effect}

\begin{figure} 
\includegraphics{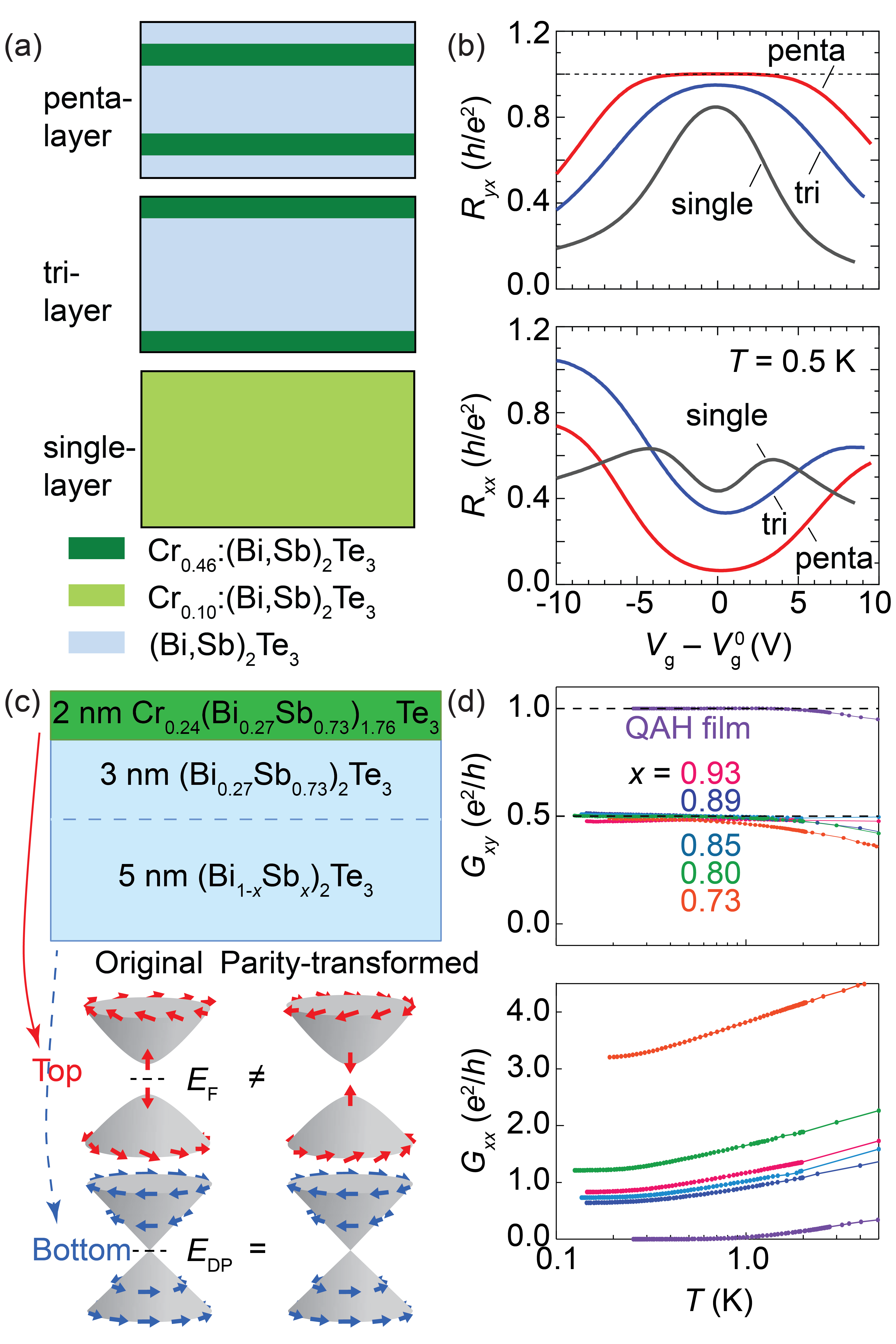}
\mycaption{\label{fig:fig4}\textbf{Magnetic modulation doping for higher quantization temperature and parity anomaly.} (a) Schematics of lightly doped uniform single-layer (bottom), modulation-doped tri- (middle) and penta-layer (top) by alternating heavily Cr-doped and undoped (Bi,Sb)$_2$Te$_3$. (b) The gate dependence of the Hall resistance $R_{yx}$ and longitudinal resistance $R_{xx}$ measured at $T$ = 0.5 K in the absence of magnetic field. (c) Schematics of asymmetric magnetic TI with gapped top and gapless bottom surface Dirac states, enabling condensed matter investigation of the relativistic parity anomaly. (d) The half-integer quantization of the zero-field Hall conductance $G_{xy}$ and the accompanying sheet conductance $G_{xx}$. Adapted with permission from ref. \ignorecitefornumbering{\cite{RN111}}, American Institute of Physics, (a-b); ref. \ignorecitefornumbering{\cite{RN117}}, Springer Nature Limited, (c-d).}
\end{figure}

In early QAH studies, the critical temperature $T\textsubscript{QAH}$ reaching full quantization is rather low in the mK range. By Cr/V codoping, the $T\textsubscript{QAH}$ can be enhanced to 0.3 K with $R_{yx}(0)$ = $h/e^2$ (within the experimental uncertainty) and $R_{xx}(0)$ = 0.009 $h/e^2$, while at $T$ = 1.5 K, $R_{yx}(0)$ = 0.97 $h/e^2$ and $R_{xx}(0)$ = 0.19 $h/e^2$ \cite{RN104}. The increase in the $T\textsubscript{QAH}$
 via Cr/V codoping verifies to have originated from the improved magnetic homogeneity and favorably modulated surface band structure. 

Further increase in the $T_{\textrm{QAH}}$ would be beneficial in order to advance quantitative understanding of the QAH physics and to facilitate practical applications. There are apparent disadvantages in introducing ferromagnetic order via traditional doping though: ($i$) The foreign species act as defects that inevitably degrade the sample quality; ($ii$) The innate random distribution of magnetic dopants lead to undesirable disorder (including spin scattering) and fluctuation in the energy band (in addition to the more generic local variations of Bi/Sb ratio and/or film thickness) that adversely affect the QAH state \cite{RN105,RN106,RN107,RN108,RN109}; ($iii$) At the high level of doping, or rather substitution, needed for enhanced $T_{\textrm{C}}$, the band structure is prone to the development of impurity states in the bulk gap and its topological nature may also be affected, let alone the likelihood of secondary phase segregation \cite{RN110}. Thus, unfortunately, the doping route comes with inherent limitations. 

The challenging yet highly desirable goal of raising the $T\textsubscript{QAH}$ is being actively pursued. A successful approach involves the strategy of magnetic modulation doping. Specifically, as shown in Fig.~\ref{fig:fig4}(a), apart from the conventional single-layer Cr$_x$(Bi,Sb)$_{2-x}$Te$_3$ with uniform and modest Cr doping ($x$ = 0.10), alternating combinations of heavily doped ($x$ = 0.46) and pristine ($x$ = 0) TI QLs are rationally designed to form tri- or penta-layer stacks. The penta-layer structure enables an impressively high $T\textsubscript{QAH}$ = 0.5 K, while $R_{yx}(0)$ reaches 0.97 $h/e^2$ at 2 K in Fig.~\ref{fig:fig4}(b) \cite{RN111}. The greatly suppressed disorder and magnetic fluctuation are believed to be the key in enhancing the $T\textsubscript{QAH}$. In the modulation doping case, as we shall see below, there is also the possibility that the internal interfacial exchange interaction becomes very effective leading to a better exchange gap opening in the ‘cleaner/pristine’ layers of TI, and thus enabling to achieve a higher $T\textsubscript{QAH}$.

\begin{figure} 
\includegraphics{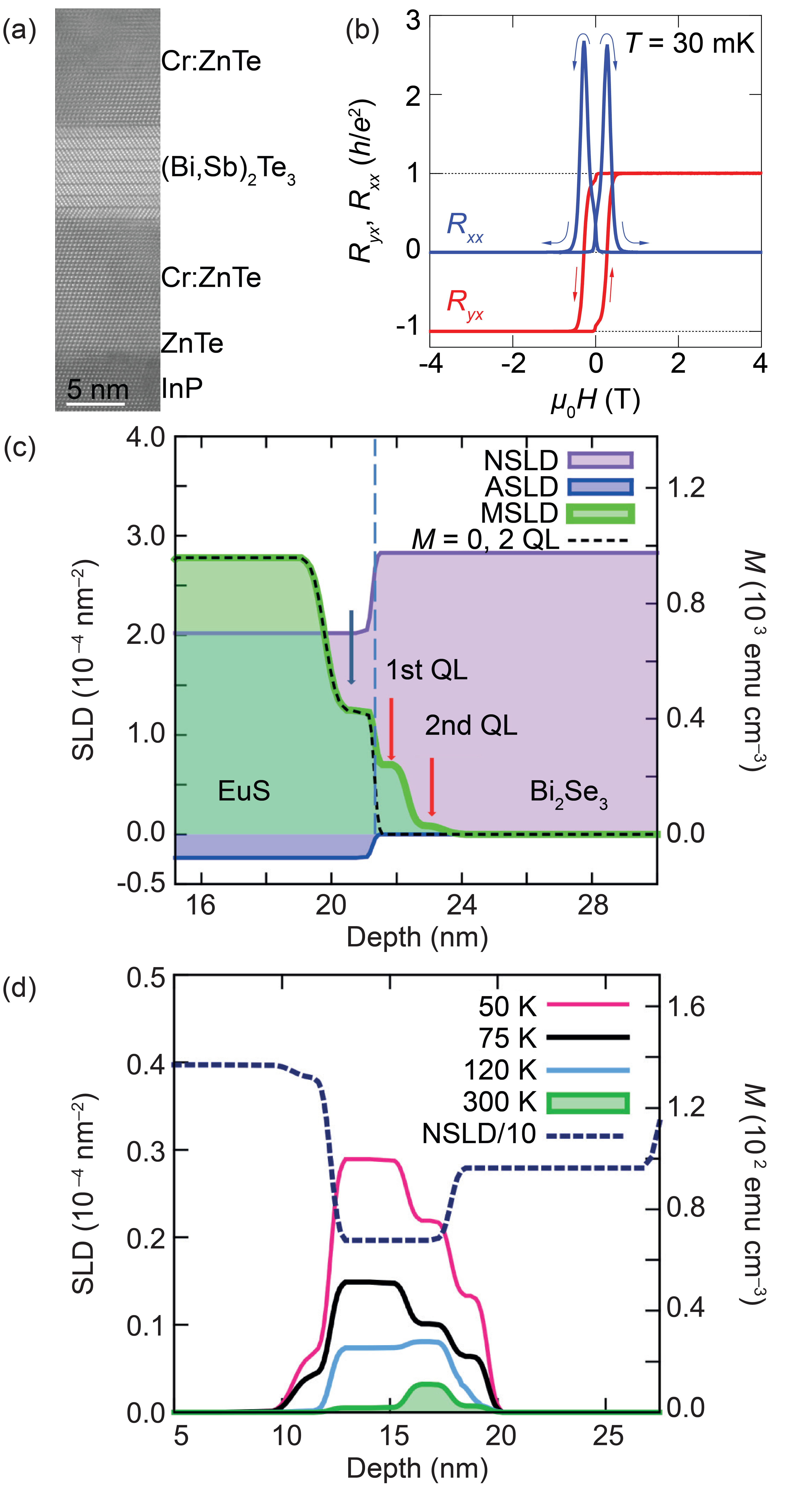}
\mycaption{\label{fig:fig5}\textbf{Proximity-driven magnetic coupling at magnetic-topological insulator interface.} (a) A cross sectional high-angle annular dark-field scanning transmission electron microscopy (HAADF-STEM) image of a Cr:ZnTe / (Bi,Sb)$_2$Te$_3$ / Cr:ZnTe stack. (b) The realized QAH state at $T$ = 30 mK. (c) The PNR nuclear (NSLD), absorption (ASLD) and magnetic (MSLD) scattering length density profiles as a function of the distance from the sample surface, measured for 5 nm EuS/20 QL Bi$_2$Se$_3$ sample at 5 K under applied in-plane magnetic field of 1 T. The magnetization measured inside the Bi$_2$Se$_3$ layer is marked with red arrows, and the reduction of the in-plane component of EuS at the interface caused by a canting of the Eu magnetization vector towards the out-of-plane direction is marked with a blue arrow. The fit with zero magnetization ($M$ = 0 in 2 QL) in the Bi$_2$Se$_3$ layer has a large deviation from the experimental data. (d) Chemical (NSLD, dashed line) and magnetic (MSLD) depth profiles at a few selected temperatures (solid lines for 50 K, 75 K and 120 K and green shading for 300 K). The scale on the right shows the magnetization. Adapted with permissions from ref. \ignorecitefornumbering{\cite{RN118}}, American Institute of Physics, (a-b); ref. \ignorecitefornumbering{\cite{RN122}}, Springer Nature Limited, (c-d).}
\end{figure}

It is worth pointing out that, at such high level of Cr substituion on the Bi/Sb sites (exceeding 20\%), the SOC is expected to be much weakened and the band inversion may no longer sustain. It hence drives the Cr-rich layer away from the TI phase towards a magnetic insulator (MI) instead. The versatility of the MI-like heavyly Cr-doped TI block has been recently recognized and extended to an architecture of [3 QL Cr:(Bi,Sb)$_2$Te$_3$ / 4 QL (Bi,Sb)$_2$Te$_3$]$_n$ / 3 QL Cr:(Bi,Sb)$_2$Te$_3$ \cite{RN112}. Similarly to an earlier TI-normal insulator design of [6 QL (Cr,V):(Bi,Sb)$_2$Te$_3$ / 3.5 nm CdSe]$_n$ / 6 QL (Cr,V):(Bi,Sb)$_2$Te$_3$ multilayers \cite{RN113}, it successfully facilitates the demonstration of the high Chern number (adjustable by the stacking number $n$) QAH state \cite{RN114,RN115}. This design also allows for probing the physics underlying the plateau-to-plateau phase transition connecting the Chern number $C$ = 1 and $C$ = 2 phases, by means of systematic engineering the strength of SOC, via tuning the Cr concentration, in the middle Cr-doped layer of a penta-layer device \cite{RN116}. The Chern number tunable QAH state attests to the capability and precise control of multichannel dissipationless chiral conduction. Recently, as illustrated in Fig.~\ref{fig:fig4}(c), asymmetric placement of the Cr-rich block in a semi-magnetic configuration allows selective opening of an exchange gap only in the top surface of TI \cite{RN117}. It enables condensed matter investigation of the relativistic parity anomaly in quantum field theory, offering long-sought experimental verification of the half-integer quantization of $G_{xy}$ in Fig.~\ref{fig:fig4}(d).

\section{Proximitized MI/TI interface}

The revelation of the role of MI/TI interfaces in modulation doped TI points to the significance of proximitized internal exchange coupling. Indeed, the QAH effect has been recently achieved in MI/TI/MI heterostructures \cite{RN118}. As shown in Fig.~\ref{fig:fig5}(a), Cr-doped ZnTe, a large gap MI with favorable lattice parameters in the (111) plane matching TI, can be grown with non-magnetic TI into high quality sandwiches. The stacks possess atomically sharp interfaces and minimal Cr interfacial migration, as confirmed by energy-dispersive X-ray spectroscopy (EDS) mapping and distribution profile analysis \cite{RN118,RN119,RN120}. Full quantization has been observed at $T$ = 30 mK, see Fig.~\ref{fig:fig5}(b). Despite the high growth temperature, the insignificant Cr diffusion across such MI/TI interface is a salient feature, enabling reliable placement of alternating Cr-doped TI and pristine TI layers, instrumental for realizing the higher Chern number QAH state \cite{RN112}, as well as spin-orbit-torque-based electrical switching of the chirality of edge current \cite{RN121}.  

The proximity-driven engineering offers a promising avenue for further enhancing the $T_\textrm{QAH}$ beyond the presently available sub-Kelvin regime as well as uncovering new physics. Higher temperature QAH states may be obtained by imprinting magnetism via high-quality heterostructure with compatible MI possessing high $T_\textrm{C}$ (or N\'eel temperature $T_\textrm{N}$ for antiferromagnet), while in the meantime preserving the structural integrity and salient surface electronic properties of TI. \emph{The interface approach is a center piece at the forefront of the field.} 

\begin{figure*} 
\includegraphics{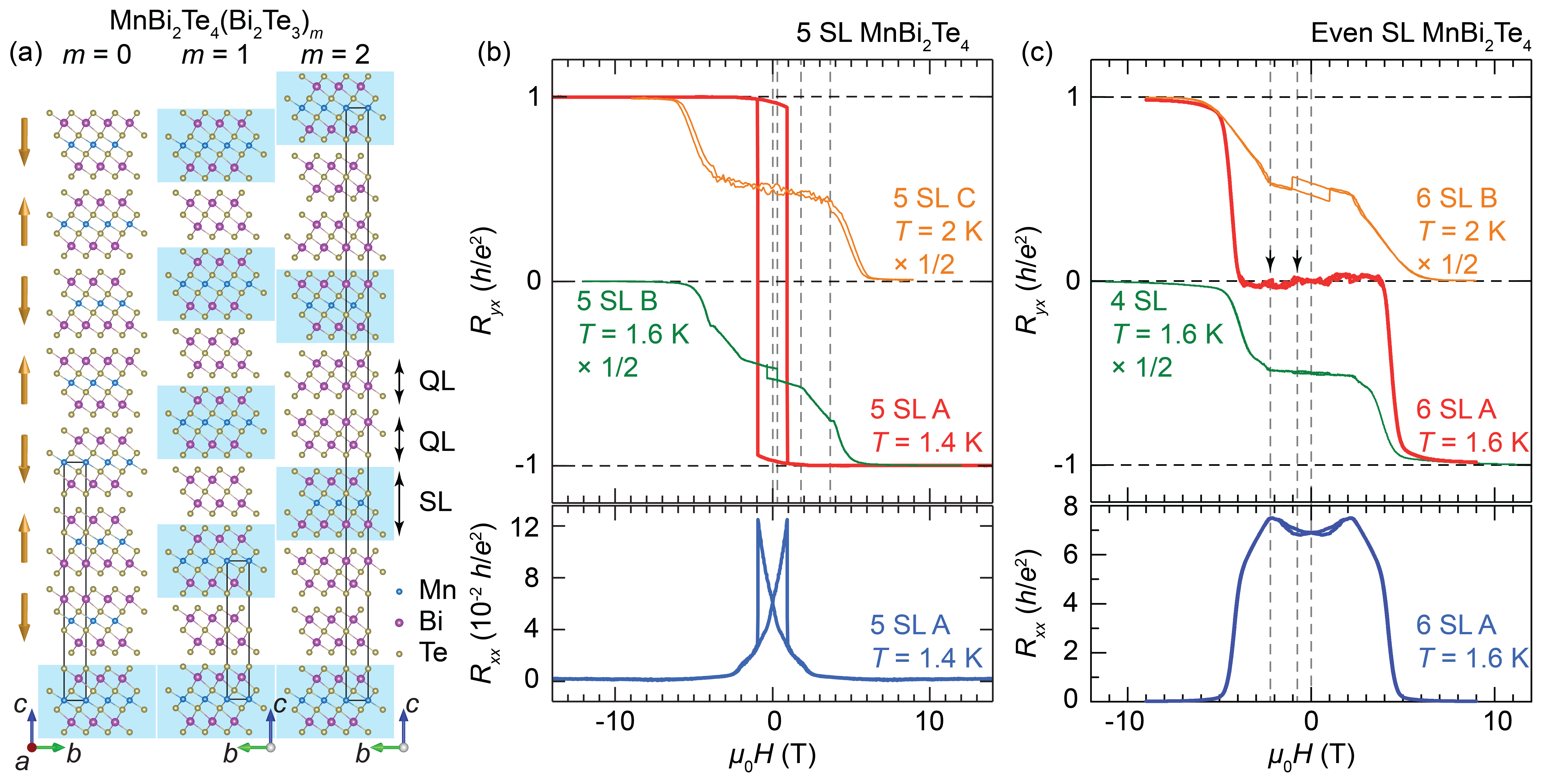}
\mycaption{\label{fig:fig6}\textbf{Intrinsic magnetic topological insulator MnBi$_2$Te$_4$.} (a) Crystal structure of natural superlattice of BiMn$_2$Te$_4$(Bi$_2$Te$_3$)$_m$, with $m$ = 0, 1, 2, ... The A-type antiferromagnetic coupling of the septuple layers (SLs) of MnBi$_2$Te$_4$ may be tuned by intercalating quintuple layers (QLs) of non-magnetic Bi$_2$Te$_3$. Field dependence of $R_{yx}$ (top) and $R_{xx}$ (bottom) for the (b) quantum anomalous Hall and (c) axion insulator like state in optimized 5 SL (A) and 6 SL (A) MnBi$_2$Te$_4$, respectively. Additional 5 SL (B and C) devices and even layer samples, 6 SL (B) and 4 SL, are shown for comparison (scaled by $\times$1/2 and vertically shifted for clarity). Adapted with permissions from ref. \ignorecitefornumbering{\cite{RN62}}, American Association for the Advancement of Science, (b) 5 SL A, 5 SL B, (c) 4 SL; ref. \ignorecitefornumbering{\cite{RN160}} Springer Nature Limited, (c) 6 SL A; ref. \ignorecitefornumbering{\cite{RN157}} American Chemical Society, (b) 5 SL C, (c) 6 SL B.}
\end{figure*}

Recently, polarized neutron reflectometry (PNR) experiments at EuS/Bi$_2$Se$_3$ interface have demonstrated that long-range magnetic order can be induced at the surface of TI without the complication of creating spin-scattering centers in the doping case \cite{RN122}. As shown in Fig.~\ref{fig:fig5}(c), the magnetic scattering length density (MSLD) profile reveals ferromagnetism extends $\sim$ 2 nm into the Bi$_2$Se$_3$ at 5 K. The experimental spin asymmetry ratio, defined as SA = $(R^{+} - R^{-})/(R^{+} + R^{-})$ with $R^{+}$ or $R^{-}$ being the reflectivity with neutron spin parallel ($+$) or antiparallel ($-$) to the external field, provides sensitive measurement of in-plane magnetism (SA = 0 designates no magnetic moment). The temperature evolution of SA profiles in EuS film and EuS/Bi$_2$Se$_3$ bilayer are drastically distinct \cite{RN122} -- while SA becomes vanishingly small in EuS above $T\textsubscript{C}$ $\sim$ 17 K, magnetism apparently survives up to 300 K in EuS/Bi$_2$Se$_3$ interface, see Fig.~\ref{fig:fig5}(d) for MSLD depth profiles at selected temperatures. This enhanced magnetic behavior has been further confirmed by superconducting quantum interference device (SQUID) magnetic measurements as well as X-ray magnetic circular dichroism (XMCD) studies. It is worth emphasizing that particularly clean, sharp and controlled interface between EuS and TI is required in achieving the needed effective magnetic proximity coupling due to the extreme short-range nature of the exchange interaction ($\lesssim$ 0.5 nm) \cite{RN123,RN124,RN125,RN126}. Extraordinary care is warranted in the TI growth, and the magnetic interfacial hetero-structuring thereafter, to properly mitigate the adverse influence from residual chalcogen atoms since optimal TI demands a Se/Te-rich growth condition. This intriguing discovery has opened a vibrant arena for proximitized MI/TI heterostructures \cite{RN127,RN128,RN129,RN130,RN131,RN132,RN133,RN134,RN135,RN136,RN137,RN138}.

Extensive investigations have since been devoted to interfacial exchange coupling and the magnetic proximity effect in MI/TI systems \cite{RN139} using Y$_3$Fe$_5$O$_{12}$ (YIG) \cite{RN140,RN141}, Tm$_3$Fe$_5$O$_{12}$ (TIG) \cite{RN142}, MnTe \cite{RN143}, Cr$_2$Ge$_2$Te$_6$ \cite{RN119,RN144} and Fe$_3$GeTe$_2$ \cite{RN145}, just to name a few. Convincing out-of-plane hysteretic AH data in platforms utilizing MI with in-plane bulk magnetic anisotropy are not yet available and should be pursued. Initial studies leveraging TIG, a high-$T\textsubscript{C}$ ($\sim$ 560 K) MI with perpendicular magnetic anisotropy, have shown the ordering of TI can be increased to $> 400$ K \cite{RN142}. The $R_{yx}$ at present is admittedly small though, on the order of 0.7 -- 2 $\Omega$ at 2 K. Inspired by the high $T\textsubscript{C}$, it warrants future effort to improve the interfacial coupling, for instance, by growing TIG and TI all \emph{in situ} without breaking vacuum during the interface fabrication. Sandwich heterostructures taking advantages of multiple interfaces \cite{RN118,RN119} such as TIG/TI/TIG are also desirable, should one be able to control the magnetic anisotropy of TIG as out-of-plane oriented in subsequent layers. Furthermore, proximitizing magnetic TI with antiferromagnetic (Al:)Cr$_2$O$_3$ enables exchange biasing at the interface \cite{RN146,RN147}, allowing for further tunability in QAH-based devices \cite{RN148}.

\section{Intrinsic magnetic TI – MnBi$_2$Te$_4$}

\begin{figure*} 
\includegraphics{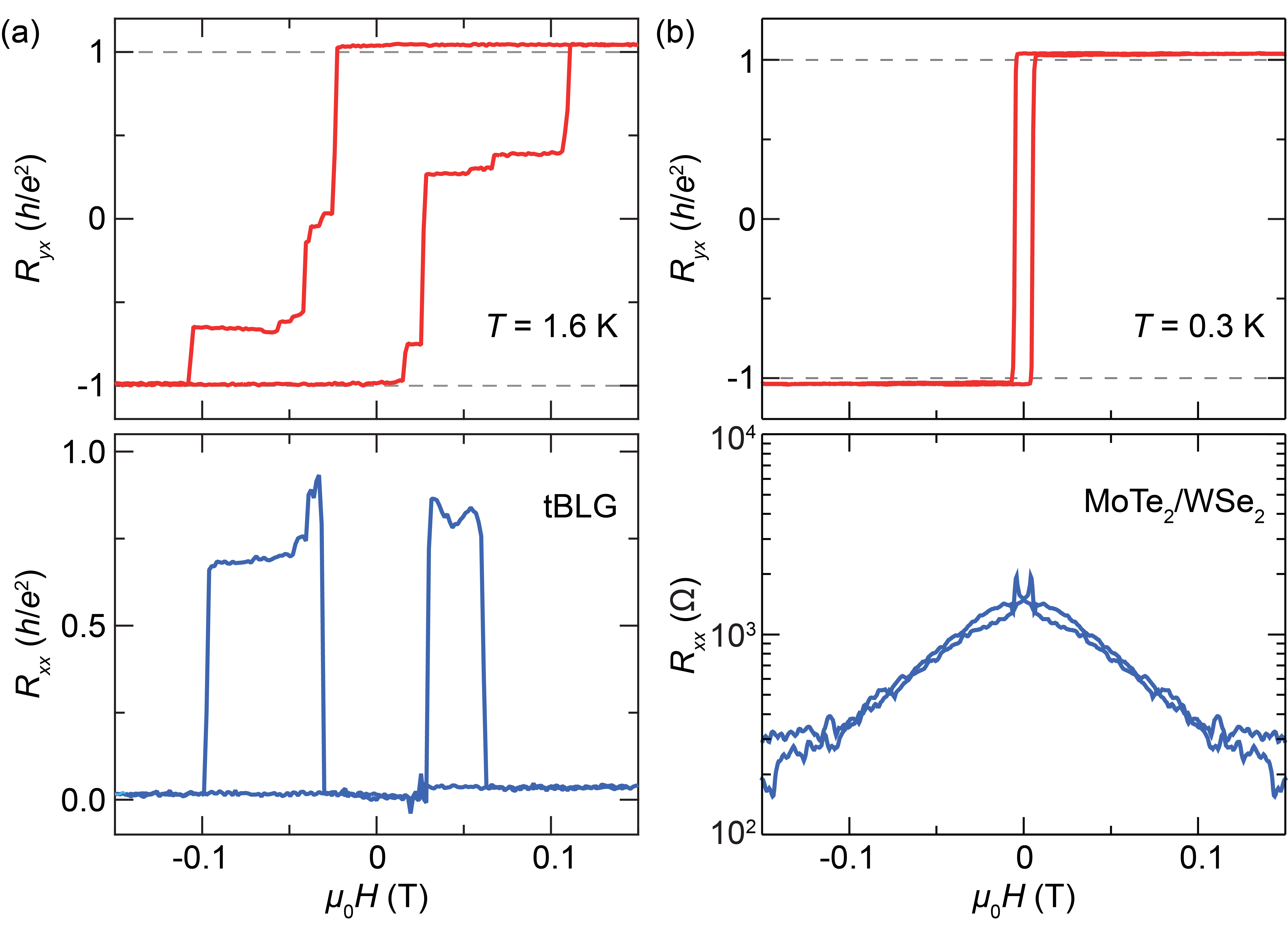}
\mycaption{\label{fig:fig7}\textbf{Quantum anomalous Hall effect in moir\'e materials.} Field dependence of $R_{yx}$ (top) and $R_{xx}$ (bottom) for (a) magic-angle twisted bilayer graphene (tBLG) and (b) AB-stacked MoTe$_2$/WSe$_2$ heterobilayers. Adapted with permissions from ref. \ignorecitefornumbering{\cite{RN63}}, American Association for the Advancement of Science, (a); ref. \ignorecitefornumbering{\cite{RN64}} Springer Nature Limited, (b).}
\end{figure*}

Recognizing the critical importance of high-quality magnetic interfaces in QAH related physics, one may envision to push the proximity-driven phenomena down to its ultimate limit – i.e. taking advantage of the layer-by-layer atomically ordered internal interfaces in intrinsic magnetic TIs. The promise and feasibility of such strategy were perhaps well hinted by the large magnetic gap seen in natural superlattice of QL Bi$_2$Te$_3$ and septuple layer (SL) MnBi$_2$Te$_4$ in Mn-doped Bi$_2$Te$_3$ \cite{RN149}. As shown in Fig.~\ref{fig:fig6}(a), MnBi$_2$Te$_4$ SLs feature intralayer ferromagnetic order and A-type interlayer antiferromagnetism. The interlayer coupling is tunable via intercalation of non-magnetic Bi$_2$Te$_3$ QLs, forming a family of MnBi$_2$Te$_4$(Bi$_2$Te$_3$)$_m$ with rich topological and physical properties \cite{RN150,RN151,RN152,RN153,RN154,RN155}. 

The layer sensitive magnetism in MnBi$_2$Te$_4$ offers a versatile platform exploring the magnetic field driven transitions connecting the (canted) antiferromagnetic and poled ferromagnetic phases in the few SL regime, as well as the rich QAH (odd) and/or axion (even) insulator physics with precise SL number control \cite{RN156,RN157,RN158}. As illustrated in Fig.~\ref{fig:fig6}(b), the QAH effect has been realized in one optimal 5 SL MnBi$_2$Te$_4$ (device 5 SL A), displaying $R_{yx}(0)$ = 0.97 $h/e^2$ and $R_{xx}(0)$ $<$ 0.061 $h/e^2$ at 1.4 K \cite{RN62}. While the $T_{\textrm{QAH}}$ for full zero-field quantization was not reported, it is likely on par with the state-of-the-art thin film results using Cr/V codoping \cite{RN104} and Cr-modulation doping \cite{RN111}. Comparable QAH state has also been achieved in crystal flakes with MnBi$_2$Te$_4$/Bi$_2$Te$_3$ natural superlattice \cite{RN159}. It has stimulated intense investigations in intrinsic magnetic TIs, although reproducing the stunning QAH result \cite{RN62} in 5 SL MnBi$_2$Te$_4$ still appears to be challenging to date. It is likely owing to complex interplay among crystal quality, magnetic/lattice defect, stacking/strain condition, fabrication processing, device morphology and substrate/film/capping interfaces, which leads to often qualitatively different $R_{yx}(H)$ profiles, as shown in Fig.~\ref{fig:fig6}(b), e.g., for devices from different batch (device 5 SL B) \cite{RN62} or different group (device 5 SL C) \cite{RN157}. 

As depicted in Fig.~\ref{fig:fig6}(c), a zero Hall plateau has been observed in 6 SL MnBi$_2$Te$_4$ (device 6 SL A) \cite{RN160}, establishing another candidate system for solid state investigation of axion physics \cite{RN161,RN162,RN163}. The characteristic critical fields, corresponding to the small but discernible kinks in $R_{yx}$ and extrema in $R_{xx}$ for device 6 SL A \cite{RN160}, are generally consistent with other even SL devices (6 SL B \cite{RN157} and 4 SL \cite{RN62}). By means of electric field tuning, Berry-curvature-induced layer Hall (LH) effect has been uncovered in this axion insulator like regime \cite{RN164}. The capability of fusing magnetism and topology at the atomic level catalyzes future advancement of topological quantum effects. To further understand the key factors underlying the magnetic and topological nature of MnBi$_2$Te$_4$, and QAH insulators in general, multimodal investigations are of immediate interest, synergizing magnetotransport and various scanning probes including magnetic force microscopy (MFM) \cite{RN165}, microwave impedance microscopy (MIM) \cite{RN166}, nano-SQUID \cite{RN107}, magneto-optical Kerr effect (MOKE) and magnetic circular dichroism (MCD) \cite{RN167}. 

\section{moir\'e materials}
By interfacing 2D crystal layers, of either the same specie at a small twist angle or different kind possessing dissimilar lattice parameters, one creates artificial moir\'e superlattices hosting intertwined topology and strong correlations \cite{RN168}. Moir\'e graphene heterostructures with valley-spin-degenerate topological flat bands enable remarkable phenomena including superconductivity \cite{RN169}, correlated insulating states \cite{RN170}, and when TRS is broken, orbital magnetism featuring moir\'e-scale current loops \cite{RN171}. Significant AH effect with strong magnetic hysteresis has been observed in twisted bilayer graphene (tBLG) \cite{RN172} and ABC-trilayer graphene/hexagonal boron nitride (ABC-TLG/hBN) moir\'e superlattices \cite{RN173}. As shown in Fig.~\ref{fig:fig7}(a), a robust QAH state has been demonstrated at 1.6 K in a narrow range of density near band filling factor $\nu$ = 3 in a flat-band tBLG device aligned to hBN \cite{RN63}. $R_{yx}$ quantization, within 0.1\% of $R_{\textrm{K}}$, has been found to survive up to 3 K, while the systems displays a $T_{\textrm{C}}$ of 7.5~K. 

In electrically tunable semiconductor-based heterobilayers, the application of an out-of-plane gating electric field modulates the bandwidth as well as the band topology by intertwining moir\'e bands from different layers. At $\nu$ = 1, the QAH effect has been achieved at 0.3 K, upon spontaneous valley polarization in MoTe$_2$/WSe$_2$ moir\'e superlattice with AB configuration \cite{RN64}, see Fig.~\ref{fig:fig7}(b). $R_{yx}$ remains quantized up to about 2.5 K, while staying finite up to $T_{\textrm{C}} \sim 5-6$ K. The two newly emerged moir\'e platforms manifest $R_{yx}$ $>$ $h/e^2$ when approaching quantization, different from the TI-based phenomenology with $R_{yx}$ $<$ $h/e^2$, hinting different disorder mechanisms might underpin the QAH effect owing to orbital magnetic states. 

\section{Conclusion and outlook}
In recent years, tremendous progress has been made advancing topological concepts in solid state. The realization and development of the QAH effect attest to the ever-more coherent synergy of theory prediction and interpretation as well as experimental exploration and discovery of new materials. We expect the fundamental interfacial magnetism in heterostructures to fuel future development of the field. Identifying new magnetic topological systems with suitable properties for implementing the QAH effect, or exploring the interplay between magnetism and topology in general, is of paramount interest \cite{RN174,RN175,RN176}. The versatile selective tunability of magnetic topological interfaces further enables investigation of high energy physics counterparts in materials laboratory, such as dyon particles \cite{RN177} and Majorana bound states \cite{RN178,RN179} when additional superconductivity proximity is coupled. 

From an application standpoint, the QAH effect bodes particularly well for future universal quantum units standard combining the Josephson effect in one measurement setup, where uncertainty in the 1 ppb range is prerequisite to rival that of existing QH systems \cite{RN180}. The ideally dissipation-free nature of chiral edge transport in the QAH state inspires low-energy consumption spintronics or integration to existing computing architectures as chiral interconnects \cite{RN181}. Furthermore, when hybridized with superconductors, QAH insulators motivate topological quantum computing \cite{RN182}. These technological breakthroughs leverage on the capability of interfacial engineering structural, chemical, magnetic and electronic properties at the atomic scale, preferably with robust QAH state for practical operation temperature. 

Despite impressive progress using interface-inspired approaches (heterostructure or intrinsic), the $T_{\textrm{QAH}}$ to manifest quantized $R_{yx}$ with negligible $R_{xx}$ ($\ll$ 1\% $h/e^2$) is still limited in the sub-Kelvin regime for all Bi$_2$Te$_3$-derived QAH systems. It implies mechanisms intrinsic to tetradymites, e.g., spatial inhomogeneities in thickness and/or prevailing antisite defects, might be the main culprit limiting the operation temperature, in lieu of the magnetic ordering as $T_{\textrm{C}}$/$T_{\textrm{N}}$ is reasonably high on the order of tens of Kelvin. Indeed, recent analysis of non-local transport in a Corbino geometry has revealed QAH edge channels surviving up to $T_{\textrm{C}}$ in V-doped TI \cite{RN183}. Further optimizing the bulk insulation in tetradymites \cite{RN184}, or discovering entirely new class of TIs beyond the current paradigm \cite{RN185}, is beneficial towards increasing the practical relevance of the QAH effect. 

In additional to MBE growth, recent development in alternative and industrial friendly route such as sputtering \cite{RN186} may bring about previously overlooked benefits including quantum confinement and/or superior sample quality. The dynamic MBE deposition and equilibrium crystal synthesis bear dramatically different kinetics and thermodynamics. To realize the QAH state, the former route is instrumental in preparing doped films beyond the bulk solubility limit, while the latter is critical in ensuring the needed layer ordering and placement in intrinsic TIs, however not typically vice versa. Realizing intrinsic magnetic TI films capable of quantization is highly desirable, meaning materials exploration. Prospects of advancing QAH physics also lie in the exploration of possibly fractional QAH by feasibly introducing the needed strong correlation, as well as novel mechanisms exploiting e.g., in-plane magnetization \cite{RN187} and antiferromagnetism \cite{RN188}. Further inspiration can also be drawn from the exciting development of twistronics benefiting from magnetic states of the orbital origin \cite{RN189}.

\section*{Acknowledgments}
The work was supported by Army Research Office (W911NF-20-2-0061), National Science Foundation (NSF-DMR 1700137 and CIQM 1231319) and Office of Naval Research (N00014-20-1-2306). H.C. was sponsored by the Army Research Laboratory under Cooperative Agreement Number W911NF-19-2-0015.

\nocite{*}
\bibliography{0-MS}

\end{document}